\newcommand{\paperauthors}{\IEEEauthorblockN{Mahdi Banisharif\IEEEauthorrefmark{1},
Tabassum Mahmud\IEEEauthorrefmark{1},
Ahmad Tarraf\IEEEauthorrefmark{2},
Suren Byna\IEEEauthorrefmark{3},
Ali Jannesari\IEEEauthorrefmark{1}}
\IEEEauthorblockA{\IEEEauthorrefmark{1}Iowa State University, USA,
\{msharif, tmahmud, jannesar\}@iastate.edu}
\IEEEauthorblockA{\IEEEauthorrefmark{2}Technical University of Darmstadt, Germany,
ahmad.tarraf@tu-darmstadt.de}
\IEEEauthorblockA{\IEEEauthorrefmark{3}The Ohio State University, USA,
byna.1@osu.edu}}
\newcommand{\paperpdfauthor}{Mahdi Banisharif, Tabassum Mahmud, Ahmad Tarraf, Suren Byna, Ali Jannesari}
\newcommand{\paperpdftitle}{IOSage: Detecting HPC I/O Bottleneck Patterns to Guide LLM Recommendations}
\newcommand{\referencesbreak}{}

\newcommand{\acknowledgmentsbreak}{\clearpage}
\documentclass[conference]{IEEEtran}
\usepackage[T1]{fontenc}
\IEEEoverridecommandlockouts

\usepackage{cite}
\usepackage{amsmath,amssymb,amsfonts}
\usepackage{algorithmic}
\usepackage{graphicx}
\usepackage{textcomp}
\usepackage{xcolor}

\usepackage{booktabs}       %
\usepackage{array}          %
\usepackage{multirow}       %
\usepackage{xspace}         %
\usepackage{placeins}       %
\usepackage{enumitem}       %
\usepackage{microtype}      %
\usepackage{hyperref}
\makeatletter\g@addto@macro\UrlBreaks{\do\a\do\b\do\c\do\d\do\e\do\f\do\g\do\h\do\i\do\j\do\k\do\l\do\m\do\n\do\o\do\p\do\q\do\r\do\s\do\t\do\u\do\v\do\w\do\x\do\y\do\z\do\0\do\1\do\2\do\3\do\4\do\5\do\6\do\7\do\8\do\9}\makeatother

\graphicspath{{figures/}{./}}

\providecommand{\paperauthors}{\IEEEauthorblockN{Anonymous Authors}
\IEEEauthorblockA{Anonymous Affiliations}}
\providecommand{\paperpdfauthor}{}
\providecommand{\paperpdftitle}{}
\providecommand{\referencesbreak}{\clearpage}

\providecommand{\acknowledgmentsbreak}{}

\hypersetup{
    colorlinks=true,
    linkcolor=black,
    citecolor=black,
    urlcolor=blue,
    pdfauthor={\paperpdfauthor},  %
    pdftitle={\paperpdftitle},
}

\newcommand{\system}{IOSage\xspace}       %

\newcommand{\pGran}{access granularity\xspace}      %
\newcommand{\pMeta}{metadata intensity\xspace}      %
\newcommand{\pImb}{parallelism efficiency\xspace}   %
\newcommand{\pSeq}{access pattern\xspace}           %
\newcommand{\pColl}{interface choice\xspace}        %
\newcommand{\pFiles}{file strategy\xspace}          %
\newcommand{\pSync}{throughput utilization\xspace}  %

\newcommand{\newnum}[1]{#1}  %

\def\BibTeX{{\rm B\kern-.05em{\sc i\kern-.025em b}\kern-.08em
    T\kern-.1667em\lower.7ex\hbox{E}\kern-.125emX}}

\begin{document}

\title{IOSage: Detecting HPC I/O Bottleneck Patterns to Guide LLM Recommendations}

\author{\paperauthors}

\maketitle

\begin{abstract}
HPC applications often leave much of their parallel file system's bandwidth
unused, and turning their Darshan logs into concrete I/O fixes still requires
a scarce I/O expert.
Rule-based tools flag counters against thresholds set by experts, and in
tools built on large language models (LLMs), such as ION and IOAgent, the LLM
decides which bottlenecks a log contains.
We present \system{}, in which a classifier trained on many labeled logs, not
the LLM, makes that decision, since a single log gives an LLM no baseline for
what is abnormal.
The LLM receives the I/O bottleneck patterns the classifier detects, the
job's configuration files, per-file Darshan evidence, and for each pattern a
card built from benchmark training runs, and writes recommended changes.
Applied by hand and measured in repeated paired runs, the changes
\system{} names speed up median total wall time \newnum{4.38}$\times$ for an
openPMD kernel (\newnum{4.02}$\times$ on a second storage tier) and
\newnum{1.42}$\times$ for hourly WRF restarts.
For \newnum{5} of the \newnum{6} application scenarios with a helpful change, \system{} with Claude
Sonnet~4 names the helpful change, including the change
for 10-minute WRF restarts, where it speeds up median total wall time
up to \newnum{1.93}$\times$; with GPT-4o and Llama~3.1 as well, the
classifier and cards name more helpful changes than the LLM alone.
On \newnum{78} benchmark test runs without a pattern, the LLM deciding alone
proposes changes for \newnum{64}; with the classifier, for \newnum{9}.
The classifier reaches \newnum{0.929} Micro-F1 on \newnum{488} labeled benchmark
test runs (\newnum{0.916} as the gate; \newnum{0.832} on the \newnum{158} from configurations and jobs
outside training), above Drishti's rules, WisIO, and IOAgent's pipeline
without its literature retrieval.
On TraceBench's \newnum{nine} expert-labeled traces from production systems,
it has the highest precision of the tools compared (\newnum{0.82}), with
\newnum{2} false positives against \newnum{12} for the keyword-mapped LLM
diagnoses shipped with TraceBench and \newnum{5} for Drishti, on the five
issue labels that match registered patterns.

\end{abstract}


\begin{figure*}[t]
  \centering
  \includegraphics[width=\textwidth]{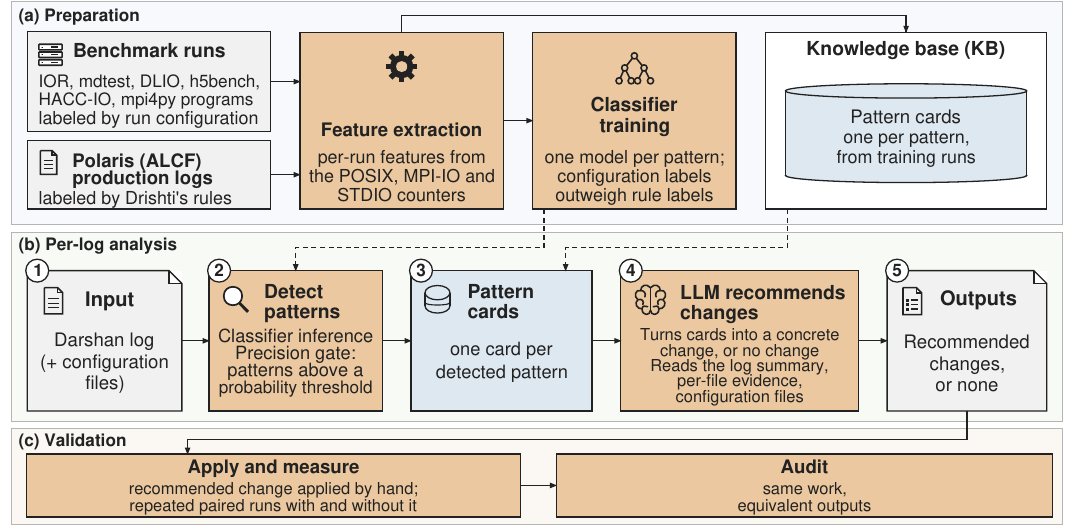}
  \caption{\system{} architecture. (a)~Preparation: configuration-labeled
  benchmark runs and rule-labeled Polaris logs train one classifier per
  pattern; the training runs yield one pattern card per pattern (typical
  counter values and a change template). (b)~Per-log analysis: the LLM's
  prompt holds the detected patterns with their cards, the log's summary and
  per-file evidence, and the job's configuration files. (c)~Validation: a
  recommended change is applied by hand and its wall-time effect measured.
  Grey: inputs and outputs; orange: computation; blue: stored products;
  dashed arrows: products of (a) used in (b).}
  \label{fig:architecture}
\end{figure*}

\section{Introduction}
\label{sec:intro}

Applications on high-performance computing (HPC) systems often use only a
fraction of the available I/O bandwidth~\cite{luu2015multiplatform,
bez2023survey}.
On Polaris at the Argonne Leadership Computing Facility
(ALCF)~\cite{alcf2025polaris}, \newnum{131{,}151} of a collection of
\newnum{1.40} million Darshan~\cite{snyder2016darshan} logs meet our cleaning
criteria (Section~\ref{sec:background}), and in about \newnum{46\%} of them
the \system{} classifier detects at
least one bottleneck pattern, such as small transfer sizes, random access on a
parallel file system, or independent instead of collective MPI-IO.
These logs hold \newnum{97.8\%} of the I/O time Darshan records, summed over
ranks.

Recognizing the patterns requires an I/O specialist who can read Darshan
counters, reason about Lustre striping and MPI-IO hints, and relate the
counters to application code.
Few application teams have such a specialist~\cite{egersdoerfer2025ioagent}.

Existing tools follow three approaches.
Rule-based tools such as Drishti~\cite{bez2022drishti} flag metrics that
cross thresholds set by experts, and WisIO~\cite{yildirim2025wisio} maps
metrics to bottlenecks with an extensible rule set.
They are fast and their decisions are easy to trace, and their thresholds
and rules are written by experts rather than learned from logs.
Learned approaches such as AIIO~\cite{dong2023aiio} train models on Darshan
counters and show which counters explain slow I/O, but they predict
throughput, not which patterns are present, and they do not recommend changes.
Tools built on large language models (LLMs), such as
ION~\cite{egersdoerfer2024ion} and its successor
IOAgent~\cite{egersdoerfer2025ioagent}, which adds retrieval over 66 I/O
papers, explain the I/O behavior in a Darshan log in natural language; in
both, the LLM decides which patterns a log contains.
On our labeled benchmark test runs (Section~\ref{sec:eval}), most patterns
that IOAgent's diagnosis pipeline, run from its public code on the log alone,
like the classifier, without its literature retrieval (IOAgent without
retrieval below), reports with
GPT-4o~\cite{openai2024gpt4o} are absent from the labels: its precision (the share of reported patterns that
are present) is \newnum{0.218} under our mapping of its text to the labels.
STELLAR~\cite{egersdoerfer2025stellar} has LLM agents read Darshan counters,
tune Lustre parameters for the workload, measure each attempt, and keep the
rules they learn for later workloads; it tunes the file system for a
workload, while \system{} decides which patterns the log of a finished job
shows and recommends changes to the application's I/O.

Deciding which patterns a log shows needs a baseline of what is normal, and
a single log does not contain one.
A classifier trained on many labeled logs learns that baseline from data.
\system{} therefore gives the pattern decision to a classifier, used as a
precision gate: it decides which patterns the prompt names, and the LLM turns those patterns into concrete changes
(Fig.~\ref{fig:architecture}).
Our method is this division of labor and its measurement: removing each
part in turn shows what it adds, on applications and on benchmark runs with
known labels.

We label benchmark runs by the configuration that induces each pattern, so
detection is not scored against a rule over the counters the model reads,
and keep part of them for evaluation; the multi-label classifier trains on
the rest together with counter-derived labels of production logs, which
receive a lower weight (biquality weighting).
In \system{}, the prompt names the registered patterns (those the
classifier learns to detect) that the classifier finds in the log, with a card for each from a
knowledge base (KB) built from benchmark training runs; the prompt also
holds the job's configuration files and per-file Darshan evidence.
The configuration files and per-file evidence let \system{} tailor a
recommendation to an application without reading its source code.

A card shows how a pattern is usually removed, not whether a change helps a
given application, so on applications the changes \system{} names are
applied by hand and judged by total wall time over repeated runs.

Each card gives a pattern's counter signature and a generic change template;
the LLM turns that template into a concrete change for the application, such as switching the I/O library for
WRF~\cite{skamarock2019wrf} restart files.
This paper makes three contributions:
\begin{itemize}[leftmargin=*]
  \item \textbf{A learned pattern decision before the LLM's
    recommendations.} A multi-label classifier selects the registered
    patterns and their pattern cards, and the LLM uses these, per-file
    Darshan evidence, and the job's configuration files to propose changes
    (Section~\ref{sec:design}); on \newnum{78} benchmark test runs without a
    pattern, Claude Sonnet~4 proposes a change for \newnum{64} when deciding
    alone and for \newnum{9} within \system{} (Section~\ref{sec:eval:rec}).
  \item \textbf{Configuration-labeled detection with production data.}
    Benchmark runs are labeled by the configuration that induces each
    pattern and combined with rule-labeled production logs at a lower
    weight; the classifier reaches \newnum{0.929} Micro-F1 (the harmonic
    mean of precision and recall pooled over all labels, from 0 to 1, higher
    better) on \newnum{488} labeled test runs, \newnum{0.916} as the gate, and
    \newnum{0.832} on the
    \newnum{158} from configurations and jobs outside training
    (Section~\ref{sec:eval:detection}).
  \item \textbf{Measured outcomes on applications.} We apply the changes
    \system{} names by hand and compare repeated, paired runs by total wall
    time (Section~\ref{sec:eval:apps}): with Claude Sonnet~4, \system{}
    names a helpful change for \newnum{5} of the \newnum{6} evaluated
    scenarios that have one, the applied changes give
    \newnum{4.38}$\times$ for an openPMD~\cite{poeschel2021openpmd} kernel
    and \newnum{1.42}$\times$ to \newnum{1.93}$\times$ for WRF restarts, and
    a change the LLM proposes alone slows Nek5000.
\end{itemize}

Section~\ref{sec:background} describes the I/O behavior on Polaris that
motivates the design, and Sections~\ref{sec:related}
and~\ref{sec:limitations} cover related work and limitations.

\section{Background and Motivation}
\label{sec:background}

\subsection{What Darshan Logs Record}
\label{sec:background:darshan}
Darshan~\cite{snyder2016darshan} is deployed at several leadership
facilities, including ALCF, where it produced the logs we use.
It writes one log per execution of an MPI application, and one log per
process for non-MPI programs.
For each file, the POSIX module keeps operation counts, byte totals,
access-size histograms, the four most frequent access sizes and strides,
timestamps of the first and last operations, and cumulative read, write, and
metadata time.
The MPI-IO module keeps similar counters without strides or sequential-access
counts, and it separates independent from collective operations.
The STDIO module keeps counts, bytes, timestamps, and times but no histograms or
access sizes.

These counters show how an application issued its I/O, for example many
small writes or independent MPI-IO on a shared file.
They do not show why: the log holds no application semantics, no state of the
storage system or of other jobs, and only coarse timing within a file.
Monitoring that captures only the I/O interfaces cannot fully explain
bottlenecks inside deeper layers of the I/O stack~\cite{neuwirth2025xio}.
Tracers such as DFTracer~\cite{devarajan2024dftracer} record events across
layers, but they are opt-in and cannot be applied to existing log archives.
Time-resolved behavior, such as periodic I/O phases~\cite{tarraf2024ftio,
boito2025temporal}, is also largely invisible in aggregate counters.
The HEATMAP module records I/O volume over time.
\system{} does not use HEATMAP data.
We therefore treat each bottleneck pattern as I/O behavior that the counters
of a Darshan log can show, not as a diagnosed cause.

\subsection{Production I/O on Polaris}
\label{sec:dataset:data}
Our production data is the ALCF Polaris Darshan log
collection~\cite{alcf2025polaris}, which is publicly available and holds
\newnum{1{,}397{,}216} anonymized logs from April 2024 to February 2026.
Polaris has \newnum{560} HPE Apollo nodes, each with one AMD EPYC CPU and four
NVIDIA A100 GPUs.
Its Lustre file systems include Eagle, with \newnum{160} object storage targets
(OSTs), \newnum{40} metadata targets (MDTs), \newnum{100}\,PB of capacity, and
\newnum{650}\,GB/s of peak bandwidth.
Of all logs, only \newnum{2.9\%} record MPI-IO, and \newnum{46.4\%} move less
than 1\,MB in total.

\begin{figure}[!t]
  \centering
  \includegraphics[width=\columnwidth]{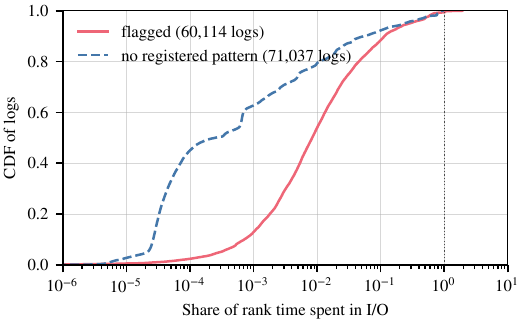}
  \caption{Share of rank time in recorded I/O calls for the \newnum{131{,}151} cleaned
  Polaris logs: POSIX and STDIO read, write, and metadata time summed over
  processes, divided by processes times run time. Values above 1 occur when
  calls overlap (\newnum{397} logs).}
  \label{fig:motivation}
\end{figure}

Cleaning keeps logs with POSIX records, at least 1\,s of run time, at least
1\,KB and one POSIX read or write, and non-negative Darshan timers, which
leaves \newnum{131{,}151} logs.
The Polaris cohort and the production training logs therefore hold no
STDIO-only log; the gate still scores such a log at inference
(Section~\ref{sec:design:inference}).
Logs with a registered pattern spend more of their rank time in I/O
(Fig.~\ref{fig:motivation}): \newnum{6{,}952} of the \newnum{60{,}114}
flagged logs spend at least 10\%, and the median share is \newnum{0.83\%} for
flagged logs against \newnum{0.025\%} for the others.

\subsection{Challenges}
\label{sec:background:problems}
\noindent\textbf{1) No ground truth in production logs.}
Production logs carry no pattern labels, so labels for them are computed
from the log itself.
Independent labels come from benchmark runs whose configuration induces the
pattern: an IOR run with 64-byte transfers carries the \pGran{} label because of its
configuration, not because of a threshold on the counters in its log.
Such labels describe benchmark behavior.
Inadequate dataset coverage is a known error source of I/O
models~\cite{isakov2022taxonomy}, so generalization to applications must be
tested separately.

\noindent\textbf{2) One log is not a baseline.}
An LLM that reads a Darshan log judges it alone, with neither the thresholds
of a rule tool nor the distribution of logs on the system; in our evaluation,
LLMs deciding alone report patterns absent from the labels and propose
changes for runs without a pattern (Section~\ref{sec:eval}).
A classifier trained on many labeled logs learns from data which counter
values mark each pattern.

\noindent\textbf{3) A citation is not a measured effect.}
Whether a change helps can be shown only by running it, and on a shared
parallel file system, run-to-run I/O noise~\cite{isakov2022taxonomy} makes
one run with and one without the change insufficient.

\section{Design}
\label{sec:design}

\system{} takes a Darshan log that the facility already collects; for an
application, the user may also supply the run's input configuration files.
It works in detection and recommendation (Fig.~\ref{fig:architecture}b,
steps 2 to 5) and validation (Fig.~\ref{fig:architecture}c), with a
classifier and pattern cards prepared once (Fig.~\ref{fig:architecture}a).
In detection, a classifier trained on the labels of
Section~\ref{sec:design:labels} finds the bottleneck patterns of
Section~\ref{sec:design:patterns} in the log.
The classifier acts as a precision gate
(Section~\ref{sec:design:inference}).
In recommendation, each detected pattern selects its pattern card from a
knowledge base (KB, Section~\ref{sec:design:kb}) built from training runs
only, and the LLM writes changes from a prompt that names the patterns and
holds their cards, the log's evidence, and the job's configuration files
(Section~\ref{sec:method:llm}).
In validation, a recommended change is applied by hand and the application
runs with and without it (Section~\ref{sec:design:validation}).

\subsection{Bottleneck Patterns}
\label{sec:design:patterns}
\system{} detects the seven patterns of Table~\ref{tab:patterns}, each a
behavior that a known change can remove.
These are the registered patterns of Section~\ref{sec:intro}; a log with
none of them has no registered pattern.
A log can hold several at once, for example small random writes, so
detection is multi-label.
The changes act on transfer sizes, access order, the I/O interface or
library, file layout, striping, rank balance, and synchronization.
Detection is advisory: a pattern that the application needs, such as a
required file per process, is reported like any other; for an application,
the LLM sees the configuration files and may state that no change is
justified, and the user decides whether a change fits the application
(Table~\ref{tab:patterns}, last column).
A pattern is defined by the configurations that induce it, not by a fixed
threshold: for \pGran{}, the classifier learns where the counters separate
IOR runs with 64\,B to 16\,KiB transfers from runs without a pattern, such
as 4\,MiB collective writes; for \pImb{}, the inducing configurations come
from one benchmark (Section~\ref{sec:eval:detection}).

\begin{table*}[!t]
\centering
\caption{Registered bottleneck patterns, the benchmark configurations we ran
to induce each one, the change template of its pattern card, and when the
change does not apply}
\label{tab:patterns}
\small
\setlength{\tabcolsep}{4pt}
\begin{tabular}{@{}>{\raggedright\arraybackslash}p{1.9cm}>{\raggedright\arraybackslash}p{2.4cm}>{\raggedright\arraybackslash}p{4.3cm}>{\raggedright\arraybackslash}p{3.9cm}>{\raggedright\arraybackslash}p{4.3cm}@{}}
\toprule
\textbf{Pattern} & \textbf{Behavior} & \textbf{Inducing configurations} & \textbf{Card change template} & \textbf{When the change does not apply} \\
\midrule
\pGran & many small reads or writes & IOR with 64\,B to 16\,KiB transfers; DLIO with small records; h5bench with small blocks; HACC-IO with few particles per rank & increase the transfer size to at least 1\,MB; aggregate small writes; use collective I/O & the small requests carry a negligible share of the I/O time \\
\addlinespace
\pMeta & metadata time dominates & mdtest with 1,000 to 50,000 creates, cross-node stat, deep trees & reduce the file count; batch metadata operations; use shared files for small data & the files are required outputs, such as per-step files read by other tools \\
\addlinespace
\pImb & unequal I/O across ranks & mpi4py programs in which rank~0 writes 2 to 20 times more & balance I/O across ranks; use collective I/O & one rank owns the data by design, such as configuration or log output \\
\addlinespace
\pSeq & nonsequential offsets & IOR random offsets on a shared file; DLIO with shuffled files & sort offsets; access sequentially; enable data sieving & the data layout fixes the order, or sieving would read far more than needed \\
\addlinespace
\pColl & shared-file I/O without collective MPI-IO & IOR POSIX on a shared file; MPI-IO independent with collective buffering off; h5bench independent data & use collective calls (\texttt{MPI\_File\_write\_all}); enable collective buffering & ranks write large disjoint blocks, where independent I/O performs as well \\
\addlinespace
\pFiles & too many files & IOR file per process with 64 to 256 ranks; mdtest with 100 to 5,000 files per rank; HACC-IO file per process & use shared files with collective MPI-IO, or HDF5 or ADIOS2~\cite{godoy2020adios2} & downstream tools need one file per process \\
\addlinespace
\pSync & synchronization or striping that slows writes & IOR with \texttt{fsync} after every write; DLIO checkpoints; h5bench and HACC-IO on a single OST & remove unneeded \texttt{fsync}; choose the stripe count based on file size and workload & \texttt{fsync} guards durability, such as a checkpoint commit, or the file is too small to stripe \\
\bottomrule
\end{tabular}
\end{table*}

\subsection{Labels and Training}
\label{sec:design:labels}
\label{sec:dataset:gt}
\begin{figure}[!t]
\centering
\includegraphics[width=\columnwidth]{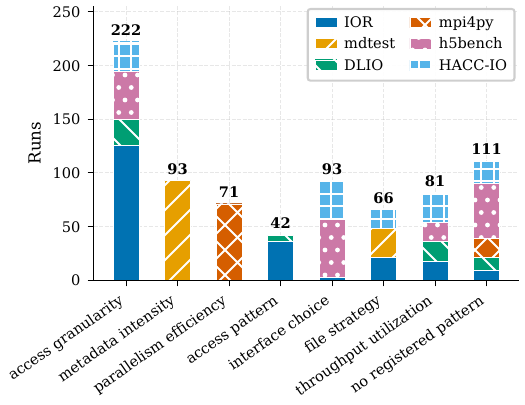}
\caption{Labels of the 689 configuration-labeled benchmark runs, by
benchmark; a run can carry several labels.}
\label{fig:labels}
\end{figure}

\noindent\textbf{Configuration labels.}
For preparation (Fig.~\ref{fig:architecture}a), we ran parameter sweeps of six benchmarks on Delta at the National
Center for Supercomputing Applications (NCSA):
IOR~\cite{shan2008ior} and its metadata companion mdtest,
DLIO~\cite{devarajan2021dlio} for deep-learning data loading,
h5bench~\cite{bez2024h5bench} for HDF5, the HACC-IO checkpoint kernel of
HACC~\cite{habib2013hacc}, and
mpi4py rank-imbalance programs.
Each run is labeled by the configuration that induces its patterns
(Table~\ref{tab:patterns}), not by its counters (Section~\ref{sec:background:problems}).
Runs of efficient configurations, such as IOR with 4\,MiB collective MPI-IO
writes, are labeled as having no registered pattern.
Fig.~\ref{fig:labels} shows the labels of the \newnum{689} runs by benchmark:
\pMeta{} comes from mdtest and \pImb{} from the mpi4py programs only, and
every other label from at least two benchmarks.

Part of the labeled runs is kept for evaluation, and the rest trains the
classifier (Section~\ref{sec:eval}).

\noindent\textbf{Production labels.}
Polaris logs are labeled by Drishti's rules~\cite{bez2022drishti}, applied
with their published thresholds to the extracted counters, and our mapping of
rule triggers to the seven patterns.
We use the \newnum{91{,}807} logs of the earliest 70\% of the
\newnum{131{,}151} cleaned logs, ordered by job start time.
These labels are plentiful, but they are fixed expert thresholds on the
counters the classifier reads.
For two patterns they also capture a symptom different from the benchmark
construction: small shared-file accesses for \pFiles, and a highest offset larger
than the bytes moved for \pSync.
This is why the configuration labels define the patterns and the
production labels add scale.

\noindent\textbf{Biquality weighting.}
Biquality learning~\cite{nodet2023biquality} trains on a small trusted label
set and a large untrusted one together.
Each configuration-labeled training run receives a weight of $w=100$ and
each production log a weight of 1, so one benchmark run counts as much as 100
production logs, while the production logs add workload diversity.
Section~\ref{sec:eval} reports how the score changes with $w$.

\noindent\textbf{Classifier.}
The classifier has one XGBoost~\cite{chen2016xgboost} model (gradient-boosted
decision trees) per label (binary relevance); tree ensembles still outperform
deep networks on typical tabular data~\cite{grinsztajn2022tree}, and other
learners can fill the same role (Section~\ref{sec:eval:detection}).
An eighth model predicts no registered pattern; Section~\ref{sec:eval:detection}
scores all eight label models at 0.5, while the gate uses the seven pattern
models at a threshold $\tau$ (Section~\ref{sec:design:inference}).
It reads \newnum{157} features computed from the POSIX, MPI-IO, and STDIO
counters of a log: operation counts, bytes, access-size histograms, times,
and derived ratios such as the share of small or
sequential accesses.
Each model has 500 trees of depth 6, learning rate 0.05, row and column
subsampling 0.8, and a per-label positive-class weight.
TreeSHAP~\cite{lundberg2020treeshap} computes SHapley Additive exPlanations
(SHAP), the contribution of each feature to one prediction, which show the
features that drive each pattern (Section~\ref{sec:eval}).

\subsection{Precision Gate}
\label{sec:design:inference}
For a new log (Fig.~\ref{fig:architecture}b, step 2), each of the seven pattern models gives a probability, and
every pattern above $\tau=0.3$ is detected.
We set $\tau$ before any evaluation, below the default of 0.5, so that a
pattern with probability between $\tau$ and 0.5 is also named in the prompt.
The gate sets which patterns and pattern cards the prompt contains
(Section~\ref{sec:design:kb}); the LLM is called for every log.
When no pattern is detected, the prompt states this, and the LLM may answer
that no change is justified.
The gate scores every log as collected, including STDIO-only logs
(Section~\ref{sec:dataset:data}).

\subsection{Knowledge Base of Pattern Cards}
\label{sec:design:kb}
The KB (step 3) holds one card per registered pattern, built from the
configuration-labeled training runs only.
A card holds the pattern's behavior, the median of each of its key counters
over the training runs that show it, where the key counters are the five
features with the largest mean absolute SHAP value for the pattern's model on
the training runs (Section~\ref{sec:eval:detection} shows the same analysis on the test
runs), the location of the inducing I/O call in one such
run's benchmark source, and the change template of Table~\ref{tab:patterns}
with code before and after.
The prompt receives the card of each detected pattern.
Cards describe patterns, not applications: they show what a pattern looks
like in counters and how it is usually removed, and the LLM derives the
concrete change from the log and the job's configuration.
\system{} uses each LLM as released, with the same cards for every model; a
new pattern adds a card.

\subsection{Recommendation}
\label{sec:method:llm}
The prompt (step 4) holds the detected patterns with their probabilities, a summary
of up to nine nonzero values of the log (for example run time, average write size, and the share
of small accesses), and the cards of the detected patterns.
For a benchmark log (the benchmark prompt), the instructions ask for a change for each detected
pattern with code before and after, and the card it follows.
For an application (the application prompt), the prompt also holds per-file evidence from the log
(the largest files with their bytes, operations, ranks, share of small
writes, and independent and collective MPI-IO writes) and the supplied
configuration files.
Its instructions ask for the smallest correct change at the level of a
configuration setting, the I/O library, or the code; the card it follows or
``none'' for each change; no invented settings or speedups; and a statement
when no change is justified.
The LLM returns JSON with a \texttt{diagnosis} field and ranked
recommendations, each with
the pattern it addresses, code or settings before and after, and its
citation.
We query Claude Sonnet~4~\cite{anthropic2026claude}, GPT-4o, and
Llama~3.1~70B~\cite{llama2024herd} at temperature 0 through
OpenRouter, a single API for several LLM vendors.

\noindent\textbf{Example.}
Fig.~\ref{fig:example} shows \system{} on a WRF run that writes restart
files every 10 simulated minutes.
The classifier detects only \pGran, and the prompt holds its card, the
per-file evidence, and WRF's \texttt{namelist.input}.
The card offers generic changes, such as aggregating small writes.
From the evidence that one rank writes each restart file through POSIX and
the setting \texttt{io\_form\_restart = 2}, Claude Sonnet~4 ranks first
switching the restart files from serial netCDF to parallel netCDF
(PnetCDF)~\cite{li2003pnetcdf}; Table~\ref{tab:apps} gives its measured effect.

\begin{figure}[!t]
\centering
\footnotesize
\fbox{\begin{minipage}{0.95\columnwidth}\raggedright
\textbf{Detected:} \pGran{} (0.9998); all other patterns below 0.01.\\
\textbf{Card:} \pGran: increase the transfer size; aggregate small writes; use collective I/O.\\
\textbf{Per-file evidence:} each restart file written by one rank, 2.8\,GB in 857 writes, 200 of them at most 1\,MiB.\\
\textbf{Configuration:} \texttt{namelist.input} (105 lines), among them \texttt{io\_form\_restart = 2}, \texttt{io\_form\_history = 2}.\\
\textbf{Explanation (abridged; the share is over the whole log):} ``68.17\% of I/O operations being small
($<$1MB transfers) \ldots{} WRF is using serial NetCDF \ldots{} where one rank
gathers data and writes through POSIX.''\\
\textbf{First:} \texttt{io\_form\_restart = 2} $\rightarrow$
\texttt{io\_form\_restart = 11}.\\
\textbf{Second:} \texttt{io\_form\_history = 2} $\rightarrow$
\texttt{io\_form\_history = 11}.
\end{minipage}}
\caption{\system{} on a WRF run that writes restart files every 10 simulated
minutes: prompt contents and Claude Sonnet~4's output (abridged).}
\label{fig:example}
\end{figure}

\subsection{Validation by Measured Wall Time}
\label{sec:design:validation}
For an application (Fig.~\ref{fig:architecture}c), we apply each change of Table~\ref{tab:apps} by hand and run the application \newnum{11} times with and \newnum{11}
without it, in round-robin order so that file-system drift affects both.
Each run without the change pairs with the run with it that follows, so the
speedup is the median of the 11 pair ratios; the interval from the 3rd to the
9th sorted ratio covers the true median ratio with a nominal \newnum{93.5\%}
probability for independent, identically distributed pairs, whatever the
shape of their
distribution~\cite{leboudec2010performance,hoefler2015}.
An audit checks that both versions do the same work and produce equivalent
outputs.
A recommendation counts when it names the applied setting, and at which
position (Section~\ref{sec:eval}).

\section{Evaluation}
\label{sec:eval}

We evaluate \system{} on three questions: do its recommendations name
changes that speed up real applications (Section~\ref{sec:eval:apps}), does
the classifier make the LLM's recommendations more precise
(Section~\ref{sec:eval:rec}), and how well does the classifier detect
patterns against independent labels (Sections~\ref{sec:eval:detection}
and~\ref{sec:eval:tracebench})?
Section~\ref{sec:eval:facility} applies the classifier to the Polaris logs.

\subsection{Setup and Metrics}
\label{sec:eval:setup}
\noindent\textbf{Platform.}
All runs use NCSA Delta CPU nodes, each with two 64-core AMD EPYC 7763
processors and 256\,GB of memory.
Applications run on one to four nodes (4 to 256 MPI ranks).
They write to \texttt{/work/hdd}, a Lustre file system on 12 disk OSTs,
except the flash-tier rows, which write to \texttt{/work/nvme}, 96 flash OSTs
of the same file system.
Total wall time is the elapsed time of the application launch
(\texttt{srun}).
We use Darshan 3.4.6, WRF 4.8.0 with PnetCDF 1.12.3, LAMMPS of 29 Aug 2024,
and openPMD-api 0.15.1.
The LLMs are queried at temperature 0, once per log and setting; temperature
0 does not make API outputs fully deterministic, so on the application cases
with the classifier each model ran four times with a fresh cache, and the
benchmark-run counts are one sample each.

\noindent\textbf{Data.}
The \newnum{689} configuration-labeled benchmark runs are split into
\newnum{201} training and \newnum{488} test runs by iterative stratification~\cite{sechidis2011stratification},
which balances the labels across the split; the test runs cover all six
benchmarks.
Most runs are kept for testing because the production logs supply most of
the training data.
The \newnum{488} test logs come from \newnum{422} jobs: some jobs write
several logs, such as the write and read phases of an h5bench job, and these
can fall on both sides of the split.
Of the \newnum{488} test runs, \newnum{158} come from configurations and
jobs outside training, and \newnum{330} share a configuration or a job with
a training run; Section~\ref{sec:eval:detection} scores both.
The pattern cards are built from the \newnum{201} training runs only.

\noindent\textbf{Metrics.}
Recall is the share of present patterns that are reported, complementing
precision and Micro-F1 (Section~\ref{sec:intro}); Macro-F1 averages the
F1 values of the seven patterns and of ``no registered pattern''.
All range from 0 to 1, with higher better.
Speedups and their intervals follow Section~\ref{sec:design:validation}; an
interval that includes 1 means no significant change, and one entirely below 1
means slower.

\subsection{Recommendations on Real Applications}
\label{sec:eval:apps}
\noindent\textbf{Measured changes.}
Table~\ref{tab:apps} lists the audited changes.
WRF writes two kinds of output: restart files, from which a run can resume,
and history files, the forecast fields; our scenarios write each once per
simulated hour (hourly) or every 10 simulated minutes (10-min) during a
one-hour simulation.
The measured changes speed up the openPMD kernel (h5bench's openPMD write
kernel, not among the labeled runs: \newnum{128} ranks write
\newnum{10.7}\,GB to one HDF5 file) \newnum{4.38}$\times$
(collective instead of independent HDF5~\cite{folk2011hdf5} transfers),
WaComM++~\cite{montella2023wacomm}
\newnum{1.95}$\times$ (no processed copy of each hourly input file), and WRF
\newnum{1.42}$\times$ to \newnum{1.93}$\times$ (parallel netCDF restart
files).
The 10-minute restart change also helps at four nodes and on the flash tier,
by less (\newnum{1.19}$\times$ and \newnum{1.34}$\times$), and parallel
netCDF history output speeds up WRF's default settings \newnum{1.17}$\times$
and the 10-minute history \newnum{1.13}$\times$.
Three changes give no gain in total wall time: the hourly history change and
two that check \system{}'s no-change answers, the shared file an LLM alone
proposes for Nek5000, which is slower, and one LAMMPS dump file per rank
(both below).
All series pass the audit: both versions do the same work and produce
equivalent outputs.
The targeted output step speeds up more than the total, up to
\newnum{5.32}$\times$ for openPMD on the flash tier and \newnum{5.18}$\times$
for WRF's default history; the hourly and 10-minute
history steps speed up \newnum{1.82}$\times$ [\newnum{1.29}, \newnum{2.34}]
and \newnum{1.44}$\times$ [\newnum{1.14}, \newnum{1.57}], but total wall time
gains only for the 10-minute output.
The restart write is faster in all six WRF restart series, the four of
Table~\ref{tab:apps} and two repeat-day series (\newnum{1.76}$\times$ to
\newnum{3.26}$\times$), across three days, two node counts, and two
storage tiers; repeated on another day, total wall time gains
\newnum{1.17}$\times$ [\newnum{1.13}, \newnum{1.30}] for the 10-minute restart
and shows no significant change for the hourly one (\newnum{1.18}$\times$
[\newnum{0.64}, \newnum{1.40}]).
We use the application's timer because Darshan's time inside I/O calls does
not fall for the WRF changes: serial netCDF gathers each field to rank~0
through MPI before writing, outside I/O calls.
Recommendations are scored against total wall time; the write-step column shows
where the time goes.

\begin{table}[!t]
\centering
\caption{Measured changes on real applications. Wall time: median of 11
runs. Speedup: median of the 11 pair ratios, with its 3rd to 9th ratio as a
93.5\% interval (Section~\ref{sec:design:validation})}
\label{tab:apps}
\footnotesize
\setlength{\tabcolsep}{1.0pt}
\begin{tabular}{@{}llrr@{\hspace{4pt}}lr@{}}
\toprule
\textbf{Application} & \textbf{Change} & \multicolumn{2}{c}{\textbf{Wall time (s)}} & \multicolumn{1}{c}{\textbf{Speedup}} & \textbf{Write} \\
\cmidrule(lr){3-4}
\textbf{and scenario} & & \textbf{without} & \textbf{with} & \multicolumn{1}{c}{\scriptsize median [3rd, 9th]} & \textbf{step} \\
\midrule
\multicolumn{6}{@{}l}{\textit{Faster: interval above 1}} \\
openPMD kernel & HDF5 collective & \newnum{76.0} & \newnum{16.6} & \newnum{4.38} {\scriptsize[2.41, 5.35]} & \newnum{4.94} \\
\quad on flash tier & & \newnum{66.1} & \newnum{16.1} & \newnum{4.02} {\scriptsize[3.68, 4.80]} & \newnum{5.32} \\
WaComM++ & no input copy & \newnum{59.1} & \newnum{30.3} & \newnum{1.95} {\scriptsize[1.47, 2.62]} & n/a \\
WRF restart, hourly & PnetCDF restart & \newnum{48.4} & \newnum{33.9} & \newnum{1.42} {\scriptsize[1.37, 1.50]} & \newnum{2.09} \\
WRF restart, 10-min & PnetCDF restart & \newnum{94.8} & \newnum{49.3} & \newnum{1.93} {\scriptsize[1.83, 1.99]} & \newnum{3.26} \\
\quad on 4 nodes & & \newnum{114.1} & \newnum{95.5} & \newnum{1.19} {\scriptsize[1.13, 1.32]} & \newnum{1.83} \\
\quad on flash tier & & \newnum{93.7} & \newnum{73.8} & \newnum{1.34} {\scriptsize[1.13, 1.61]} & \newnum{2.42} \\
WRF history, default$^{\mathrm{a}}$ & PnetCDF history & \newnum{45.0} & \newnum{38.8} & \newnum{1.17} {\scriptsize[1.12, 1.22]} & \newnum{5.18} \\
WRF history, 10-min & PnetCDF history & \newnum{52.9} & \newnum{48.7} & \newnum{1.13} {\scriptsize[1.03, 1.14]} & \newnum{1.44} \\
\addlinespace
\multicolumn{6}{@{}l}{\textit{No significant change: interval includes 1}} \\
WRF history, hourly & PnetCDF history & \newnum{39.3} & \newnum{38.6} & \newnum{0.99} {\scriptsize[0.93, 1.08]} & \newnum{1.82} \\
LAMMPS & per-rank dumps & \newnum{6.7} & \newnum{6.3} & \newnum{1.05} {\scriptsize[0.98, 1.24]} & \newnum{2.22} \\
\addlinespace
\multicolumn{6}{@{}l}{\textit{Slower: interval below 1}} \\
Nek5000 & one shared file & \newnum{30.2} & \newnum{33.5} & \newnum{0.95} {\scriptsize[0.88, 0.99]} & n/a \\
\bottomrule
\end{tabular}

\vspace{2pt}
\parbox{\columnwidth}{\footnotesize Indented rows: the change of the row above
on four nodes or on the flash tier. Hourly: one output per simulated hour;
10-min: six. $^{\mathrm{a}}$As distributed with WRF's continental-US (CONUS) benchmark case (compressed
netCDF-4 history). Write step: speedup of the targeted output step by the
application's own timer; n/a: no such timer.}
\end{table}

\noindent\textbf{Scoring.}
We use applications whose I/O changes we can build, run, and audit.
We give \system{} the log of the version without the change for the 8
scenarios on \texttt{/work/hdd} in Table~\ref{tab:apps} (without the 4-node
and flash-tier repeats and LAMMPS, below) and score whether its
recommendations name the change we applied, and at which position; 6 of them, from three applications, have a helpful change
(openPMD, WaComM++, WRF hourly and 10-minute restart, default history,
and 10-minute history).
A recommendation names the change when it contains the applied setting or
its keyword, for example \texttt{io\_form\_restart = 11}, checked by a
regular expression per scenario; each change was applied by hand and run
(Table~\ref{tab:apps}).
We developed the application prompt on these series; LAMMPS and GROMACS,
below, ran with it unchanged, and a third application's log was excluded because
its run had a CPU-binding error.

\noindent\textbf{Results.}
With pattern cards and the classifier, Claude names the helpful change for \newnum{5} of the
6 scenarios that have one and ranks \newnum{4} of them first.
For all five WRF scenarios, including the hourly history, its first-ranked
recommendation is the applied setting, whose effect Table~\ref{tab:apps} gives; for openPMD the applied
change ranks second, and for Nek5000 it recommends no change, where the change
we applied was slower (\newnum{0.95}$\times$).
Table~\ref{tab:ablation} removes the cards, the classifier, or both.
Every arm keeps the log summary, the per-file evidence, the configuration
files, and the instructions; ``no classifier'' names no pattern and holds all
seven cards, ``LLM alone'' names no pattern and holds no card, and with the
classifier a log without a detected pattern receives no card.
\system{} names the most helpful changes with every LLM.
It also proposes no change for the three logs without a detected pattern
(Nek5000, LAMMPS, and GROMACS), while without the classifier every model
proposes a change for all three; Section~\ref{sec:eval:rec} shows the same on
\newnum{78} benchmark runs without a pattern.
Two further arms, not in the table, remove one input each: without the
per-file evidence, Claude ranks the helpful change first for
\newnum{2} scenarios instead of \newnum{4}; without the configuration files,
GPT-4o names \newnum{3} instead of \newnum{5}.

\begin{table}[!t]
\centering
\caption{Helpful changes named on the 6 application scenarios (ranked first
in parentheses), by setting and LLM; one run per cell, and the \system{} row
gives the range over four runs}
\label{tab:ablation}
\footnotesize
\setlength{\tabcolsep}{4pt}
\begin{tabular}{@{}lccc@{}}
\toprule
\textbf{Setting} & \textbf{Claude} & \textbf{GPT-4o} & \textbf{Llama} \\
\midrule
\textbf{\system{}} & \textbf{\newnum{5}-\newnum{6}} (\textbf{\newnum{4}}) & \textbf{\newnum{4}-\newnum{5}} (\textbf{\newnum{4}}) & \textbf{\newnum{3}-\newnum{4}} (\textbf{\newnum{3}}) \\
no cards & \newnum{3} (\newnum{2}) & \newnum{3} (\newnum{2}) & \newnum{2} (\newnum{1}) \\
no classifier & \newnum{4} (\newnum{3}) & \newnum{3} (\newnum{2}) & \newnum{2} (\newnum{2}) \\
LLM alone & \newnum{2} (\newnum{1}) & \newnum{2} (\newnum{1}) & \newnum{1} (\newnum{0}) \\
\bottomrule
\end{tabular}

\vspace{2pt}
\parbox{\columnwidth}{\footnotesize \system{} (full prompt) in bold.}
\end{table}
For the hourly and 10-minute history scenarios, Claude ranks first the
parallel netCDF history change; it speeds up their output step
(\newnum{1.82}$\times$ and \newnum{1.44}$\times$) and total wall time only
for the 10-minute output (Table~\ref{tab:apps}).
The benchmark prompt of Section~\ref{sec:method:llm}, which holds no
per-file evidence or configuration files, leads every model to name only the
openPMD change, never first; for openPMD the classifier detects \pColl, the
pattern its helpful change addresses.
Over four fresh-cache runs, Claude ranks \newnum{4} first every time, and one
of the \newnum{12} runs, Claude's, names the WaComM++ change.

\noindent\textbf{The LLM alone.}
Here the prompt holds no detected pattern and no pattern card; it keeps
the log summary, the per-file evidence, and the configuration files.
Claude and GPT-4o then name \newnum{2} of 6, and Llama \newnum{1}.

\noindent\textbf{The classifier on logs without a pattern.}
For the Nek5000~\cite{offermans2016nek5000} fluid solver, every model deciding alone recommends a change.
Claude's is to consolidate the 32 files of each output step into one shared
file, which we measured at \newnum{0.95}$\times$, a slowdown.
With the classifier, Claude recommends no change in every setting and in
all four runs.
The same holds for LAMMPS and GROMACS.
Deciding alone, the LLM proposes changes for the LAMMPS~\cite{thompson2022lammps}
molecular dynamics code and for GROMACS~\cite{abraham2015gromacs}; for LAMMPS, all three models propose an
MPI-IO dump style that the installed LAMMPS build does not provide.
With the classifier, every model recommends no change for both.
To check whether no change is right for LAMMPS, we measured the LAMMPS option
of one dump file per rank (Table~\ref{tab:apps}).
It speeds up the output step \newnum{2.22}$\times$ (median of the pair
ratios; medians \newnum{0.82} and \newnum{0.36}\,s), while total wall time shows no significant change
(\newnum{1.05}$\times$ [\newnum{0.98}, \newnum{1.24}]).
GROMACS, which moves only about \newnum{11}\,MB through STDIO, was not
measured.
The LAMMPS log has only STDIO records, GROMACS moves no POSIX data, and
Nek5000 moves under 1\,KB through POSIX because it writes its field files
through STDIO, while every training log has POSIX data.
No pattern is detected in these logs, and \system{} with each LLM
recommends no change.

\noindent\textbf{Baselines and cost.}
Drishti (the drishti-io 0.8 package)~\cite{bez2022drishti} recommends collective MPI-IO for the slow
openPMD run.
For WRF, whose helpful change switches to parallel netCDF, and for
WaComM++, whose helpful change drops a file copy, it advises switching to
MPI-IO in general terms, without naming a setting.
Drishti checks each log against fixed thresholds and is not designed to
compare versions; for openPMD and the two WRF restart pairs (hourly and
10-minute), it reports more critical issues for the faster version than
for the slower one (\newnum{5} against \newnum{3},
\newnum{8} against \newnum{4}, \newnum{7} against \newnum{4}).
With the same three LLMs, IOAgent without retrieval~\cite{egersdoerfer2025ioagent},
which reads only the log, names collective I/O for openPMD and none of the
WRF or WaComM++ changes.
Where Drishti or IOAgent without retrieval names the change we applied, collective I/O for
openPMD, its measured effect is the openPMD row of Table~\ref{tab:apps};
their other advice for these logs names no setting to apply.
IOAgent without retrieval makes \newnum{42.5} LLM calls per log against one for \system{}, and with
Claude Sonnet~4 costs about \newnum{\$0.55} per log against \newnum{\$0.016}
for \system{} with the benchmark prompt (list prices of \$3 and \$15 per million input
and output tokens); the application prompt is two to four times
longer.

\noindent\textbf{Takeaway.}
\system{} with Claude Sonnet~4 names the parallel netCDF change for 10-minute
WRF restarts from the log and the job's configuration; there the change speeds up median total wall time up to
\newnum{1.93}$\times$.
On applications, \system{} names more helpful changes than the LLM alone
with every model (Claude \newnum{5} against \newnum{2}).
Where the classifier detects no pattern, as for Nek5000, LAMMPS, and
GROMACS, \system{} with each LLM recommends no change, while every model
deciding alone proposes one, and the one such change we measured, a shared
file for Nek5000, is slower.
Section~\ref{sec:eval:rec} measures this on benchmark runs with known
labels.

\subsection{Recommendations on Benchmark Runs}
\label{sec:eval:rec}
On the \newnum{78} test runs without a pattern, the detected patterns and
their cards set whether the LLM proposes a change.
This comparison uses the application prompt, which on benchmark runs holds
the log summary and no configuration files.
Both arms receive the same instructions, which allow the answer that no
change is justified, and differ only in the detected patterns and pattern
cards.
Deciding alone, Claude Sonnet~4 recommends a change on \newnum{64} of
these runs; each proposed change requires review and, if implemented,
repeated runs to establish its benefit (Section~\ref{sec:design:validation}).
With the classifier, Claude proposes a change on \newnum{9}; the classifier
wrongly flags \newnum{7} of these, all DLIO runs, and on the other \newnum{2}
no pattern is detected.
Where a pattern exists, the gate's cost is the patterns it does not pass to
the LLM: at $\tau=0.3$ its recall over the eight labels is \newnum{91.8\%} on the \newnum{488}
test runs, with precision \newnum{0.914}.

On all \newnum{488} test runs with the benchmark prompt
(Section~\ref{sec:method:llm}), \newnum{0.895} to \newnum{0.906} of each
model's recommendations name a pattern the run has (pattern agreement; mean
over runs with at least one recommendation; explicit no-change entries count
as none); the effect of the changes is measured in Section~\ref{sec:eval:apps}.
GPT-4o and Llama make about one recommendation per run (\newnum{1.04}) and
Claude \newnum{1.4}, and each model proposes a change on the same
\newnum{7} no-pattern runs, the runs the classifier wrongly flags.
A second setting lists all seven patterns as detected and asks for a change
for each, which shows what the LLM does when the prompt asserts every
pattern.
Claude's share then falls from \newnum{0.895} to \newnum{0.203}, its
recommendations per run rise from \newnum{1.4} to \newnum{3.7}, and it
proposes a change on all \newnum{77} no-pattern runs whose output parses.
With the classifier, every recommendation on a run with a detected pattern
follows that pattern's card.
Llama's output does not parse for \newnum{8} runs, \newnum{7} of them
without a pattern, and Claude's for \newnum{1} in the second setting.

\subsection{Detection}
\label{sec:eval:detection}
Table~\ref{tab:detection} compares the classifier with rule-based tools and
an LLM tool on the \newnum{488} test runs.
The classifier reaches \newnum{0.929} Micro-F1 against the configuration
labels (mean over five seeds, standard deviation \newnum{0.003}; 95\% bootstrap interval
\newnum{[0.912, 0.948]} for seed 42); Section~\ref{sec:eval:tracebench}
tests it against expert labels.
On the \newnum{158} test runs from configurations and jobs outside training,
it reaches \newnum{0.832} Micro-F1 (standard deviation \newnum{0.009}
over five seeds), against \newnum{0.972} on the \newnum{330} runs that share
a configuration or a job with training.
Drishti's rules, applied with their published thresholds to the extracted
counters and mapped to the patterns they signal
(Section~\ref{sec:design:labels}), reach \newnum{0.364}; they label the
production training logs, so the classifier improves on its own weak labels~\cite{ratner2020snorkel}.
For \pFiles{} and \pSync{} the rules capture a different symptom
(Section~\ref{sec:design:labels}), and Drishti scores 0 on \pFiles.
WisIO's rules~\cite{yildirim2025wisio} cover four of the seven patterns; over
all seven, where the other three count as misses, it reaches \newnum{0.311}.
IOAgent without retrieval with GPT-4o, which like the classifier receives the
target Darshan log at inference (a separate GPT-4o call maps its diagnoses to
the label set), reaches \newnum{0.331} (recall \newnum{0.688}, precision
\newnum{0.218}), at \newnum{258}\,s per log.
Two trivial baselines stay far below the classifier: per-pattern thresholds
that flag a pattern when one of one to three related features passes its
90th percentile over the production training logs (its 10th for features
where low values mark the pattern, such as the sequential-access ratios),
and the majority value of each label in those logs.
On the four patterns every tool reports, \system{} reaches \newnum{0.931}
Micro-F1, against at most \newnum{0.447} (Table~\ref{tab:detection}).
Per pattern, the classifier's F1 is \newnum{0.89} to \newnum{0.98} on these
four, while the best tool on each reaches \newnum{0.55} (Drishti and IOAgent without retrieval,
\pGran), \newnum{0.44} (Drishti, \pMeta), \newnum{0.45} (IOAgent without retrieval, \pImb), and
\newnum{0.33} (Drishti, \pSeq).
A one-sided paired bootstrap test (seed 42) finds the classifier's lead over
Drishti significant ($p<0.001$).

\begin{table}[!t]
\centering
\caption{Detection on the benchmark test runs: all eight label models at a
threshold of 0.5 (the gate, seven pattern models at $\tau=0.3$, scores 0.916)}
\label{tab:detection}
\footnotesize
\setlength{\tabcolsep}{1.6pt}
\begin{tabular}{@{}lcccc@{}}
\toprule
\textbf{Method} & \textbf{Micro-F1} & \textbf{Macro-F1} & \textbf{Hamming} & \textbf{Subset} \\
\midrule
\textbf{\system{} classifier} & \textbf{\newnum{0.929}}{\scriptsize$\pm$\newnum{0.003}} & \textbf{\newnum{0.929}}{\scriptsize$\pm$\newnum{0.004}} & \textbf{\newnum{0.019}}{\scriptsize$\pm$\newnum{0.001}} & \textbf{\newnum{0.878}}{\scriptsize$\pm$\newnum{0.006}} \\
\quad no production logs & \newnum{0.909}{\scriptsize$\pm$\newnum{0.001}} & \newnum{0.904}{\scriptsize$\pm$\newnum{0.001}} & \newnum{0.026}{\scriptsize$\pm${<}\newnum{0.001}} & \newnum{0.852}{\scriptsize$\pm$\newnum{0.003}} \\
\midrule
\multicolumn{5}{@{}l}{\textit{Other learners, same data}} \\
Random forest & \newnum{0.916}{\scriptsize$\pm$\newnum{0.003}} & \newnum{0.917}{\scriptsize$\pm$\newnum{0.005}} & \newnum{0.023}{\scriptsize$\pm$\newnum{0.001}} & \newnum{0.850}{\scriptsize$\pm$\newnum{0.008}} \\
MLP & \newnum{0.881}{\scriptsize$\pm$\newnum{0.004}} & \newnum{0.884}{\scriptsize$\pm$\newnum{0.006}} & \newnum{0.033}{\scriptsize$\pm$\newnum{0.001}} & \newnum{0.814}{\scriptsize$\pm$\newnum{0.005}} \\
\midrule
Drishti rules & \newnum{0.364} & \newnum{0.285} & \newnum{0.227} & \newnum{0.283} \\
WisIO$^{\mathrm{a}}$ & \newnum{0.311} & \newnum{0.211} & \newnum{0.310} & \newnum{0.018} \\
IOAgent, no retrieval & \newnum{0.331} & \newnum{0.284} & \newnum{0.376} & \newnum{0.052} \\
\midrule
\multicolumn{5}{@{}l}{\textit{Trivial baselines}} \\
90th percentile & \newnum{0.285} & \newnum{0.201} & \newnum{0.252} & \newnum{0.055} \\
Majority class & \newnum{0.151} & \newnum{0.034} & \newnum{0.225} & \newnum{0.160} \\
\bottomrule
\end{tabular}

\vspace{2pt}
\parbox{\columnwidth}{\footnotesize Subset: share of runs with every label
right; Hamming: share of wrong label decisions, lower better; the other
metrics range from 0 to 1, higher better. Trained models: mean $\pm$ standard
deviation over five training seeds on one split; best value per column in
bold. $^{\mathrm{a}}$WisIO covers four of the seven patterns
(\pGran, \pMeta, \pImb, \pSeq); the other three count as misses. On those four
patterns, Micro-F1 is \newnum{0.931} for \system{}, \newnum{0.447} for
Drishti, \newnum{0.425} for IOAgent without retrieval (GPT-4o), and \newnum{0.373} for WisIO.}
\end{table}

\noindent\textbf{Training choices.}
Training on the \newnum{201} benchmark runs alone gives \newnum{0.909}; the
biquality weighting adds \newnum{2.0} points.
On the same data, a random forest with the same weighting reaches
\newnum{0.916} and %
a multilayer perceptron (MLP) with standardized inputs and the same weighting
\newnum{0.881}; tree ensembles lead on this tabular data, as
expected~\cite{grinsztajn2022tree}.
With one seed, Micro-F1 rises from \newnum{0.886} at $w=1$ to between
\newnum{0.918} and \newnum{0.932} for $w$ from 10 to 500, with
\newnum{0.931} at $w=100$; every $w$ from 10 to 500 beats training on the
benchmark runs alone.
Used as the gate (seven pattern models at $\tau$, no pattern when none
fires), the classifier reaches \newnum{0.916} at $\tau=0.3$ and
\newnum{0.911} to \newnum{0.917} for $\tau$ from 0.15 to 0.5.

\begin{figure}[!t]
\centering
\includegraphics[width=\columnwidth]{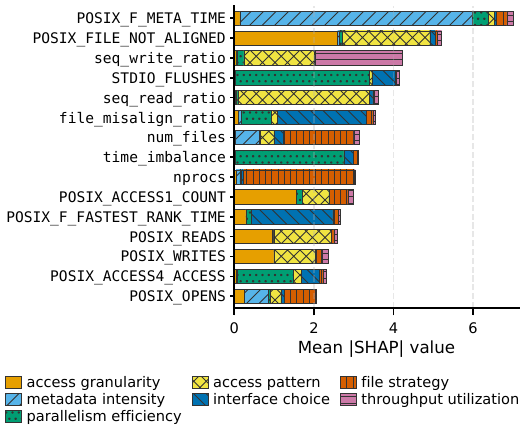}
\caption{Mean absolute SHAP value of the top counter features per pattern on
the \newnum{488} test runs (TreeSHAP on the reported classifier); each bar
segment is one pattern.}
\label{fig:shap}
\end{figure}

\noindent\textbf{Per pattern.}
Per-pattern F1 ranges from \newnum{0.81} for \pSync{} to \newnum{1.00} for
\pFiles{} (seed 42).
Fig.~\ref{fig:shap} shows the counter features that carry each pattern's
score: metadata time for \pMeta, the sequential-access ratios for \pSeq,
process and file counts for \pFiles, the unaligned-request count and the
count of the most common access size for \pGran, and file misalignment and
the fastest rank's I/O time for \pColl, the counters a practitioner checks by
hand.
For \pGran, \pColl, and \pSync{} the top feature also varies with the inducing
configuration, so Sections~\ref{sec:eval:apps} and~\ref{sec:eval:tracebench}
test the classifier outside the benchmarks, where it finds all \newnum{7}
labeled \pGran{} issues on TraceBench with no false detection and \newnum{1}
of \newnum{4} \pColl{} issues (\pSync{} has no TraceBench label).

\noindent\textbf{Per benchmark.}
Leaving one benchmark out of training (seed 42) lowers the score on its runs for
every benchmark: to between \newnum{0.26} and \newnum{0.64} for HACC-IO,
mdtest, h5bench, and IOR, from \newnum{0.41} to \newnum{0.33} for DLIO, and
to \newnum{0} for the mpi4py programs.
The classifier thus needs labeled runs of each kind of workload.

\subsection{Real-Application Traces}
\label{sec:eval:tracebench}
The public TraceBench release~\cite{egersdoerfer2025ioagent} groups \newnum{nine} traces as
real-application traces (applications, I/O kernels, and IOR and h5bench
runs), labeled by I/O experts; their logs come from five production systems
(Cori, Perlmutter, Summit, Crusher, and Theta, by their mount tables), and
the other 26 (9 single-issue programs and 17 runs of IO500, a standard storage benchmark suite) come from
test clusters, so we score the nine.
We score the five patterns that have an exact TraceBench label (\pGran,
\pMeta, \pImb, \pSeq, \pColl); misaligned
requests, which no registered pattern covers, are not scored.
We map the LLM diagnoses shipped with TraceBench (ION) to patterns by
keywords, with negations and dismissed issues removed, and use the Drishti
outputs shipped with TraceBench.
The classifier reports \newnum{2} false positives against \newnum{12} for the
LLM diagnoses and \newnum{5} for Drishti, and finds \newnum{9} true issues
against \newnum{8} and \newnum{10}; \newnum{7} of its \newnum{9} true
detections are \pGran, and both false positives are \pMeta.
Its precision (\newnum{0.82}) is the highest of the three; its F1
(\newnum{0.67}) matches Drishti's (\newnum{0.65}), a difference smaller than
one detection, and exceeds the LLM diagnoses (\newnum{0.44}); its recall
(\newnum{0.56}) is above the LLM diagnoses (\newnum{0.50}) and below Drishti
(\newnum{0.63}).
The classifier's \newnum{7} misses are \pColl{} on three traces, \pSeq{} on
two, and \pMeta{} and \pImb{} on one each.
Only one miss (\pColl{} on the SW4 seismic code, probability \newnum{0.29}) lies near $\tau$;
the other six score below \newnum{0.02}.
We accept the lower recall because a false positive turns into a proposed
change (Sections~\ref{sec:eval:apps} and~\ref{sec:eval:rec}).
With $\tau$ as set in Section~\ref{sec:design:inference}, any $\tau$ from
0.25 down to 0.10 would also catch the SW4 miss without a new false
positive; below 0.10 false positives grow.
The \pFiles{} and \pSync{} detections, which have no exact TraceBench label,
are not scored for any tool.
With misaligned requests scored, as misses for the classifier and as labels
for the tools that report them, F1 is \newnum{0.55} for the classifier,
\newnum{0.57} for the LLM diagnoses, and \newnum{0.74} for Drishti.

\subsection{Facility Study and Overhead}
\label{sec:eval:facility}
On the \newnum{131{,}151} cleaned Polaris logs, including the
\newnum{91{,}807} rule-labeled training logs, \newnum{45.8\%} show at least
one registered pattern (Fig.~\ref{fig:motivation}); the most common are \pGran{} (\newnum{30.8\%}),
\pSync{} (\newnum{20.5\%}), and \pSeq{} (\newnum{5.2\%}), and each other
pattern flags under \newnum{2\%}.
The \pSync{} and \pFiles{} shares can include the production-label symptoms
of Section~\ref{sec:design:labels}.
The flagged logs hold \newnum{97.8\%} of the I/O time Darshan records,
summed over ranks, and \newnum{81.6\%} of all rank time.
Among the \newnum{12{,}446} logs that spend at least 10\% of rank time in I/O,
\newnum{55.9\%} show a pattern, and among the \newnum{4{,}916} logs of jobs
with more than 128 processes, \newnum{83.7\%} do.
Among logs in which Darshan dropped no file records and that move at least
\newnum{100}\,GB through POSIX, those with a
registered pattern reach a median Darshan job bandwidth of \newnum{1.2}\,GiB/s,
against \newnum{2.8}\,GiB/s for the others (\newnum{1.3} against
\newnum{2.1}\,GiB/s from 1\,GB).
For one IOR log, the medians of feature extraction (10 runs) and inference
(100 runs) sum to \newnum{41}\,ms, and one LLM call in the application setting takes \newnum{3} to
\newnum{16}\,s with each LLM.

\section{Related Work}
\label{sec:related}

\noindent\textbf{Rule-based diagnosis and categorization.}
Drishti~\cite{bez2022drishti} flags Darshan metrics that cross expert
thresholds and attaches a recommendation to each trigger.
WisIO~\cite{yildirim2025wisio} computes metrics over file, process and time
views of a trace and maps them to bottlenecks and causes with an extensible
rule set.
Ather et al.~\cite{ather2024drilldown} add HDF5-level metrics and source-code
context to Darshan data.
MOSAIC~\cite{jolivel2024mosaic} segments Darshan traces with set thresholds
and assigns each job several temporal, periodic and metadata categories, to
inform job scheduling.
Each pattern in these tools is a rule that an expert writes and tunes.

\noindent\textbf{Learned models.}
Among the learned models surveyed by Peng et al.~\cite{peng2026survey},
Madhyastha and Reed~\cite{madhyastha2002classify} train a neural network on
access-pattern benchmarks, and Bez et al.~\cite{bez2019detecting} detect file
layout and access spatiality with decision trees, random forests and neural
networks.
Both classify the access pattern while the application runs, to set
file-system policy or an I/O scheduler parameter.
Gauge~\cite{isakov2020gauge} clusters the jobs of one system and fits a model
per cluster to explain throughput, and Bang et al.~\cite{bang2020clustering}
cluster workloads without labels.
AIIO~\cite{dong2023aiio} predicts a job's I/O bandwidth and explains the
prediction with SHAP, and BLG-Tuning~\cite{wang2026blgtuning} predicts I/O
time.
\system{} also learns from benchmark runs, but it assigns several bottleneck
patterns at once to the Darshan log of a finished job, each tied to known
changes, and adds rule-labeled production logs at a lower weight.
The expert-labeled traces of TraceBench~\cite{egersdoerfer2025ioagent}
serve as our external test set (Section~\ref{sec:eval:tracebench}).

\noindent\textbf{LLM-based tools.}
In ION and IOAgent (Section~\ref{sec:intro}), the LLM decides which patterns
a log shows; ION reports no quantitative detection metric, and IOAgent
evaluates its diagnoses with expert labels and an LLM judge, while we score
detection with standard multi-label metrics against configuration labels.
Metis~\cite{youssef2026metis} has LLM agents characterize the HDF5 I/O
functions themselves.
RCACopilot~\cite{chen2024rcacopilot} retrieves past cloud incidents for an
LLM, and KORAL~\cite{akewar2026koral} guides an LLM with knowledge graphs of
solid-state drive telemetry and literature; in both, the LLM decides
the root-cause category or the diagnosis.
In PG-Tune~\cite{zhang2026pgtune}, an LLM generates storage configurations
first, and a learned predictor then ranks them; in \system{}, the classifier
acts before the LLM, on what the prompt names.

\noindent\textbf{I/O tuning.}
TunIO~\cite{rajesh2024tunio} tunes I/O-stack parameters for an extracted I/O
kernel with repeated runs.
TOTO~\cite{boito2026toto} and CARAT~\cite{rashid2026carat} set Lustre stripe
counts or client request and cache parameters at run time, and
GLANCED-IO~\cite{sinurat2026glancedio} tunes deep-learning data pipelines by
greedy search on data subsets.
These tuners act while the workload runs or over repeated or reduced runs;
\system{} reads the log a job already produced and recommends changes to the
application's I/O.

\noindent\textbf{Positioning.}
\system{} differs from this work in who chooses the patterns an LLM works on
and in what the LLM draws on.
A trained classifier decides which patterns the prompt names, and the LLM
turns them into changes (Section~\ref{sec:eval:rec}).
The LLM draws only on pattern cards built from benchmark training runs and on
the target job's own log and configuration.
STELLAR~\cite{egersdoerfer2025stellar} is the closest LLM system for HPC I/O
that measures changes: it reruns the application to tune Lustre parameters,
and its learned rules carry over from benchmarks to unseen applications.
It answers which file-system settings suit a workload; \system{} answers which
patterns a finished job's log shows and which change suits the application
that wrote it (Section~\ref{sec:eval:apps}); the two are complementary.

\section{Limitations}
\label{sec:limitations}

\noindent\textbf{What a detection covers.}
\system{} reports only the seven registered patterns and recommends known
changes rather than new optimizations; adding a pattern takes a new benchmark
sweep, a change template, and retraining, and finding the root cause of a
slowdown inside application code is outside its scope.
The prompt names neither unregistered patterns, such as misaligned requests
(labeled on \newnum{six} of the nine TraceBench production-system traces),
nor patterns the classifier misses.
Section~\ref{sec:eval:tracebench} also gives a score that counts misaligned requests as misses.
\system{}'s features summarize a whole log without the HEATMAP time series,
so it cannot tell which phase or rank shows a pattern.
Different causes can also give the same detection: the classifier detects
only \pGran{} in the WRF and WaComM++ logs, whose helpful changes switch the
I/O library and drop a file copy, respectively.
The per-file evidence and configuration files lead Claude to the WRF
changes, but with the classifier only one of \newnum{12} runs names the
WaComM++ change.

\noindent\textbf{Generalization and scope.}
Micro-F1 falls to \newnum{0.832} on runs from configurations and jobs
outside training and
to between \newnum{0} and \newnum{0.64} on a benchmark left out of training;
\pImb{} has configuration labels from one benchmark only, so we do not claim
detection of rank imbalance in general (Section~\ref{sec:eval:detection}).
Training logs come from two Lustre-based systems; on the four TraceBench
traces that wrote mostly to GPFS, by their mount tables, the classifier finds \newnum{4} of the \newnum{5}
labels we score, with no false positive, and other parallel file systems are
untested.
HDF5 and netCDF reach the classifier only through POSIX and MPI-IO counters,
and the programming language does not enter the input.
The logs of LAMMPS, GROMACS, and Nek5000 carry little or no POSIX data, so
the gate's effect on logs without a pattern is measured on the \newnum{78}
of the \newnum{488} test runs that have no pattern (Section~\ref{sec:eval:rec}).
The Polaris shares are detections without ground truth.

\noindent\textbf{Recommendation evidence.}
The application results (Section~\ref{sec:eval:apps}) rest on 8 scenarios,
6 with a helpful change, and we developed the application prompt on these
series; the pattern cards come from six benchmarks.
The changes were applied by hand, not by \system{}, and two first-ranked
recommendations, source edits for openPMD and WaComM++, were not run; a cited
change may not fit another code, and we did not measure saved expert effort.

\noindent\textbf{Measurement.}
All application measurements come from one Lustre file system on Delta at
one to four nodes, and the gain of a change varies between its series: the
10-minute WRF restart change gives \newnum{1.93}$\times$ in Table~\ref{tab:apps} and
\newnum{1.17}$\times$ on a repeat day (Section~\ref{sec:eval:apps}).
Total wall time can hide a faster output step, as in the hourly WRF history
scenario (Table~\ref{tab:apps}).

\section{Conclusion}
\label{sec:conclusion}

When a trained classifier, not the LLM, decides which I/O bottleneck
patterns a Darshan log shows, the LLM proposes changes for far fewer runs that
have no pattern: on \newnum{78} such benchmark test runs, Claude Sonnet~4
deciding alone proposes changes for \newnum{64}, and with the classifier, for
\newnum{9}.
In \system{}, the LLM turns the detected patterns into changes, drawing on
pattern cards built from benchmark training runs and on the job's own log
and configuration.
Configuration labels let detection be scored against independent ground
truth, and paired runs judge each applied change by its total wall time.
Code, models, and measurements will be released on acceptance.

Next, we will apply recommendations automatically and run the remaining ones,
map detected patterns to the source lines that issue the I/O,
add per-rank and time-resolved features, compare Darshan counters with
richer traces such as DFTracer for detection, and retrain the classifier on
each facility's logs.

\acknowledgmentsbreak
\section*{Acknowledgments}
This material is based upon work supported by the National Science
Foundation under Award No.~2211982.
This work used the Delta system at the National Center for Supercomputing
Applications (award OAC 2005572) through allocation CIS240415 from the
Advanced Cyberinfrastructure Coordination Ecosystem: Services \& Support
(ACCESS) program~\cite{boerner2023access}, which is supported by National
Science Foundation grants \#2138259, \#2138286, \#2138307, \#2137603, and
\#2138296.

\referencesbreak
\bibliographystyle{IEEEtran}
\bibliography{references}

\end{document}